\documentclass[final,authoryear,3p,times,twocolumn]{elsarticle}
\usepackage{graphicx}
\usepackage{epsfig}
\usepackage{amssymb}

\journal{New Astronomy}
\begin{document}
\begin{frontmatter}
\title{The past photometric history of the FU Ori-type young eruptive star\\ 
2MASS J06593158-0405277 = V960 Mon}

\author[rj]{Rajka Jurdana-{\v S}epi{\'c}}
\author[un,um]{and Ulisse Munari}

\address[rj]{Physics Department, University of Rijeka, Radmile Matej\v{c}i\'{c}, 51000, Rijeka, Croatia}
\address[un]{corresponding author: Tel.: +39-0424-600033, Fax.:+39-0424-600023, e-mail: ulisse.munari@oapd.inaf.it}
\address[um]{INAF Astronomical Observatory of Padova, via dell'Osservatorio 8, 36012 Asiago (VI), Italy}

\begin{abstract}

The known FU Ori-type young eruptive stars are exceedingly rare (a dozen or
so confirmed objects) and 2MASS J06593158-0405277, with its 2014 outburst,
is likely the latest addition to the family.  All members have displayed
just one such eruption in their recorded history, an event lasting for
decades.  To test the FU Ori nature of 2MASS J06593158-0405277, we have
reconstructed its photometric history by measuring its brightness on Harvard
photographic plates spanning the time interval 1899-1989.  No previous large
amplitude eruption similar to that initiated in 2014 has been found, as in
bona fide FU Ori-type objects.  The median value of the brightness in
quiescence of 2MASS J06593158-0405277 is $B$=15.5, with the time interval
1935-1950 characterized by a large variability ($\sim$1 mag amplitude) that
contrasts with the remarkable photometric stability displayed at later
epochs.  The variability during 1935-1950 can either be ascribed to some T
Tau like activity of 2MASS J06593158-0405277 itself or to the also young and
fainter star 2MASS J06593168-0405224 that lies 5 arcsec to the north and
forms an unresolved pair at the astrometric scale of Harvard photographic
plates.

\end{abstract} 
\begin{keyword} Stars: pre-mail sequence
\end{keyword}

\end{frontmatter}

\section{Introduction}
\label{}

The FU Ori-type outburst of 2MASS J06593158-0405277 (hereafter `2MASS' for
short) was discovered by T.  Kojima on 3 Nov 2014 (Maehara et al.  2014). 
It has been recently given the variable star name V960 Mon (Kazarovets and
Samus 2015).  Inspection of older observations revealed the object was
already brightening during the previous two months.  No previous bright
phase of this object was known.  A low-resolution spectrum of 2MASS for 23 Nov 2014
(Maehara et al.  2014) shows no emission lines and only absorptions from
Balmer, NaI, MgI, BaII and LiI in close resemblance to that of the prototype
object FU Ori.  Hillenbrand (2014) obtained on 9 Dec 2014 a high resolution
optical spectrum, with the absorption lines showing an excellent match to
that of an early F giant or supergiant and a P-Cyg profile for H$\alpha$
indicating wind outflow.  An infrared spectrum of 2MASS was obtained on 20
Dec 2014 by Reipurth and Connelley (2014), resembling that of a late-K to early-M
star.  Such a difference in the spectral classification is typical for FU
Ori-type outbursts, with the spectral type becoming gradually later with
increasing wavelength.  This is a result of the inner warmer disk regions
dominating the optical while outer cooler disk regions dominate in the
infrared (Hartmann and Kenyon 1996).  A high spatial resolution VLT/SINFONI
infrared observation by Caratti o Garatti et al.  (2015) shows that 2MASS is
actually composed of two sources separated by 0.23 arcsec ($\sim$100 AU at
the 450 pc distance estimated by K{\'o}sp{\'a}l et al.  et al.  2015), with
a further possible and even closer third component.

FU Ori-type objects (FUORs hereafter) are pre-main sequence stars undergoing
a large amplitude outburst that typically lasts for several decades (Herbig
1977).  FU Ori itself rose in 1937 from 16.5 to 9.5 mag, where it has
remained ever since, while V1057 Cyg - that erupted in 1969 rising from 16 to
10 mag - has been steadily declining but it is still several magnitudes
brighter than its preceding quiescence (AAVSO database).  Their basic structure
has been modelled by Hartmann and Kenyon (1996).  A young, low-mass (T
Tauri) star is surrounded by a disk normally accreting at $\sim$10$^{-7}$
M$_\odot$yr$^{-1}$ onto the central star.  This slowly evolving phase is
punctuated by occasional FUOR outbursts, in which the the accretion rate
from the disk onto the star rises to $\sim$10$^{-4}$ M$_\odot$yr$^{-1}$.  The
disk becomes hot enough to radiate most of its energy at optical
wavelengths, and it dumps as much as 0.01 M$_\odot$ onto the central star
during a century-long FUOR episode.  Mass is fed to the disk by the remnant
collapsing protostellar envelope with an in-fall rate $\lesssim$10$^{-5}$
M$_\odot$yr$^{-1}$.  During FUOR eruptions, high-velocity winds
($\gtrsim$300 km sec$^{-1}$) are generated that carry away about 10\% of the
accreted material.

FUORs are rare, with about only a dozen of them being currently confirmed
according to K{\'o}sp{\'a}l et al.  (2015; the SIMBAD database lists such a
classification for a total of 39 objects).  Thus, any new addition to the
family is relevant, as for 2MASS.  The low-number statistics is still
sensitive to differences inherent to the individual objects and the
non homogeneous way they have been observed, with most of the studies focusing
on only a few FUORs.  So far, the known FUORs have displayed just one
eruption in their recorded history, but their pre-discovery photometric
behavior is sometimes undocumented.  For this reason, we went to the Harvard
photographic plate stack and directly inspected the available plates to the
aim of reconstructing the photometric history of 2MASS prior to the 2014
outburst.  The results are described in this paper.

   \begin{table}
     \caption{The photometric comparison sequence 
     around 2MASS J06593158 -0405277 used in the inspection 
     of historical Harvard plates.}
      \centering
      \includegraphics[width=85mm]{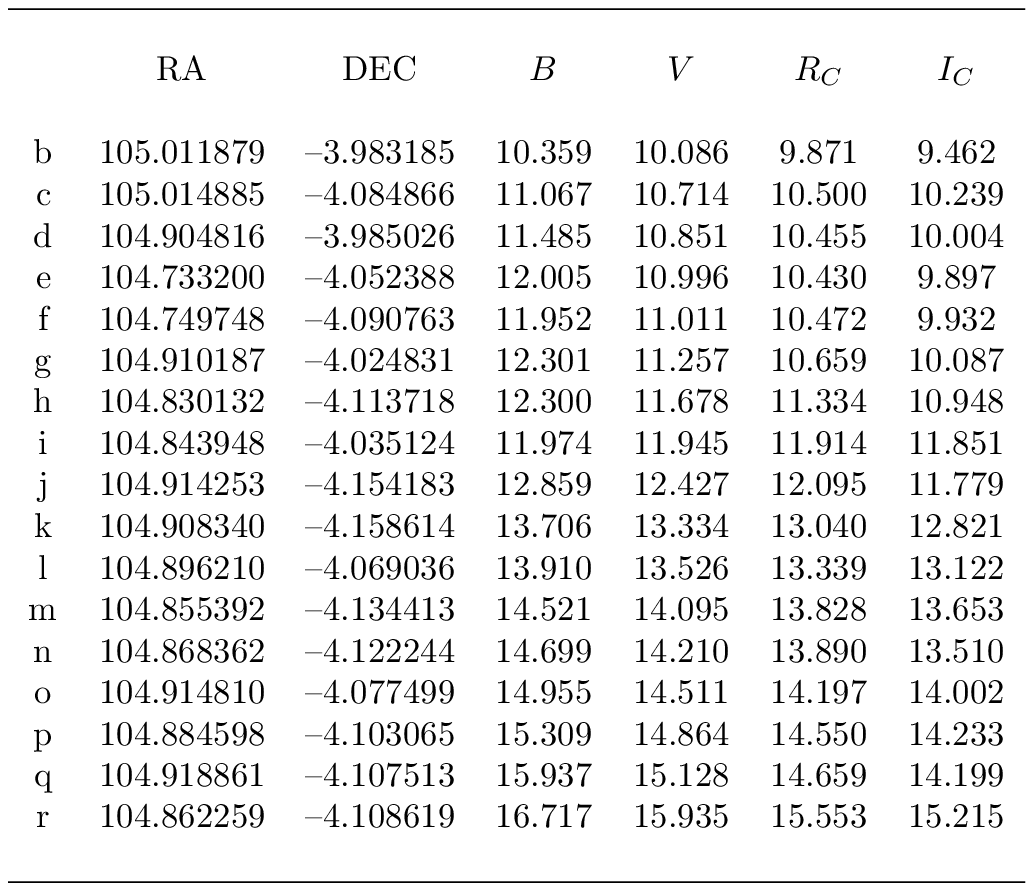}
      \label{tab1}
  \end{table}

 \begin{figure}[!Ht]
     \centering
     \includegraphics[width=8cm]{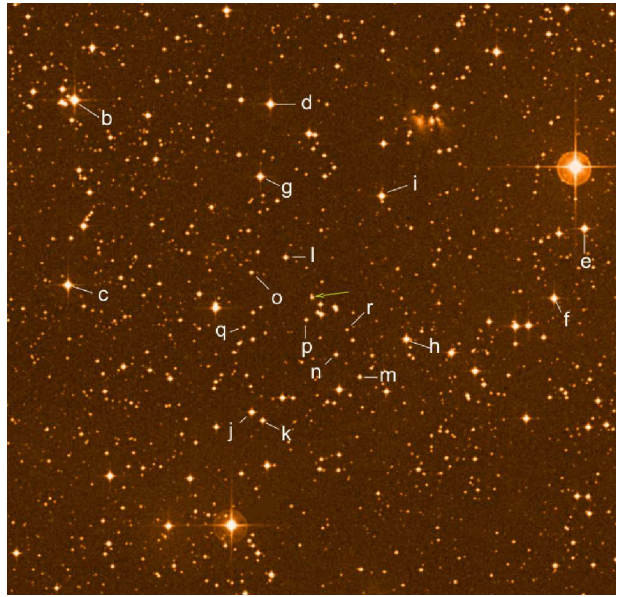}
     \caption{Identification of the photometric comparison sequence (cf. 
     Table~1) around 2MASS J06593158-0405277, placed at the center of the
     picture and identified by the arrow (north up, east to the left, size
     20 arcmin).}
     \label{fig1}
  \end{figure}

\section{Photometric sequence}

   \begin{table*}[!Ht]
     \caption{$B$ band brightness of 2MASS J06593158-0405277 on Harvard
              plates.}
      \centering
      \includegraphics[width=16cm]{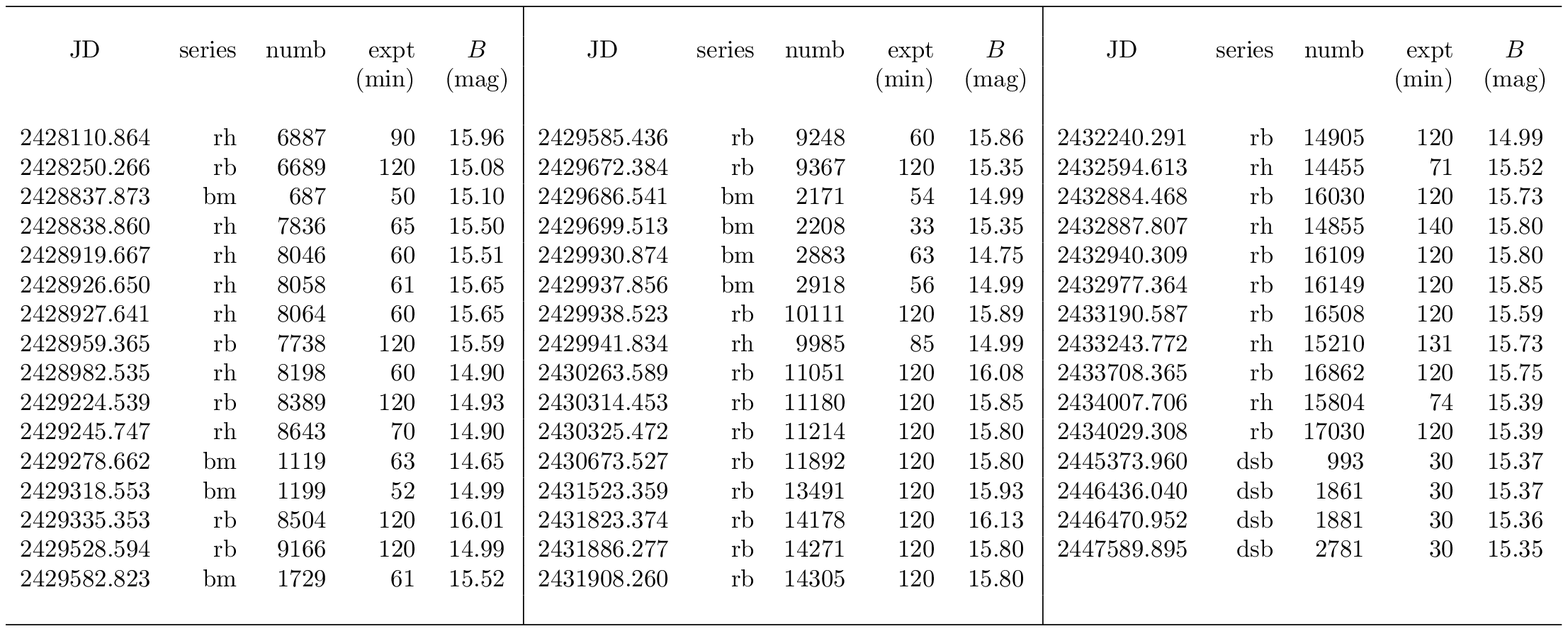}
      \label{tab2}
  \end{table*}

To estimate on Harvard plates the brightness of 2MASS, we have extracted
from the APASS all-sky photometric survey (Henden et al.  2012, Henden and
Munari 2014, Munari et al.  2014) a local photometric sequence placed around
2MASS in a way convenient for visual plate inspection.  The
sequence is presented in Table~1 and identified in Figure~1, to the aim of
assisting with the inspection of photographic plates at archives around the
world other than Harvard.

\section{Harvard plates}

Our 2MASS target has a close and fainter companion, about 5 arcsec to the
north, identified with 2MASS J06593168-0405224.  The pair is marginally
resolved on Palomar/SERC sky survey prints but it is not recognized as such
by two separate entries in the USNO-B or GSC catalogs based on them.  Both
stars show a strong excess at infrared wavelengths from circumstellar dust
(Reipurth and Connelley, 2014), and both emit in the X-rays (Pooley et al. 
2015), suggesting that also the companion is a young star as 2MASS itself. 
At the plate scale of the astrographs used to expose the plates preserved in
the Harvard stack, the pair cannot be separated and always appears as a
single, unresolved star.  Thus, our measurements of the brightness of 2MASS
on the Harvard plates actually refer to the combined pair.

 \begin{figure*}[!Ht]
     \centering
     \includegraphics[angle=270,width=15cm]{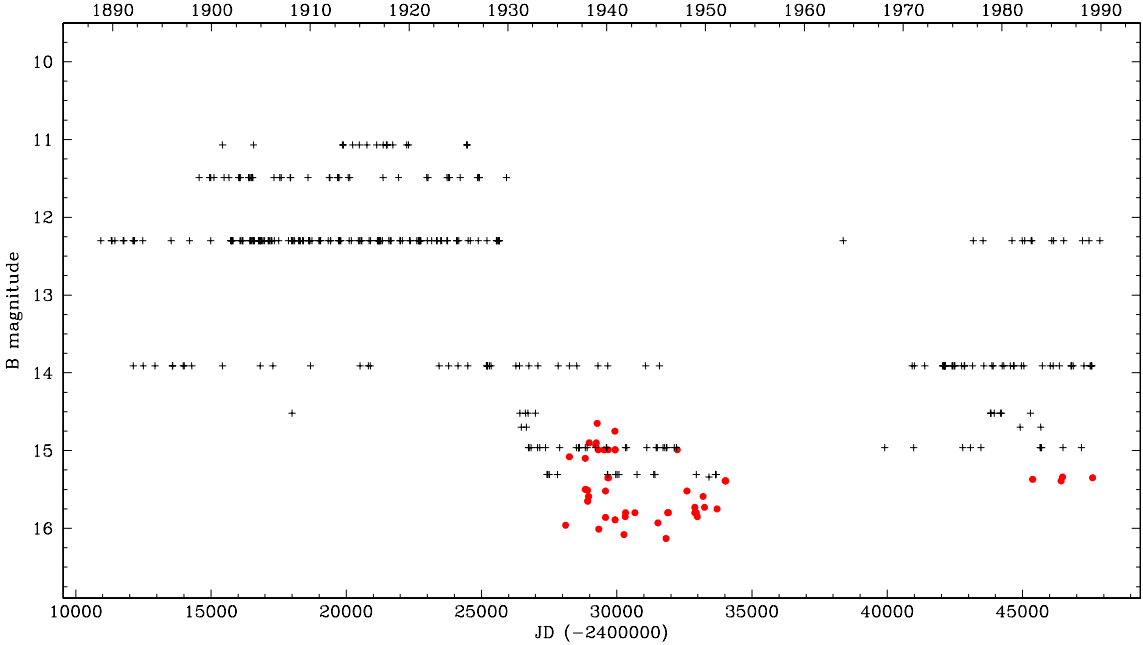}
     \caption{Historic $B$-band lightcurve of 2MASS J06593158-0405277 from Harvard
     plates. The dots mark the recorded brightness, the crosses are upper limits
     from shallower plates.}
     \label{fig2}
  \end{figure*}

 \begin{figure*}[!Ht]
     \centering
     \includegraphics[angle=270,width=15cm]{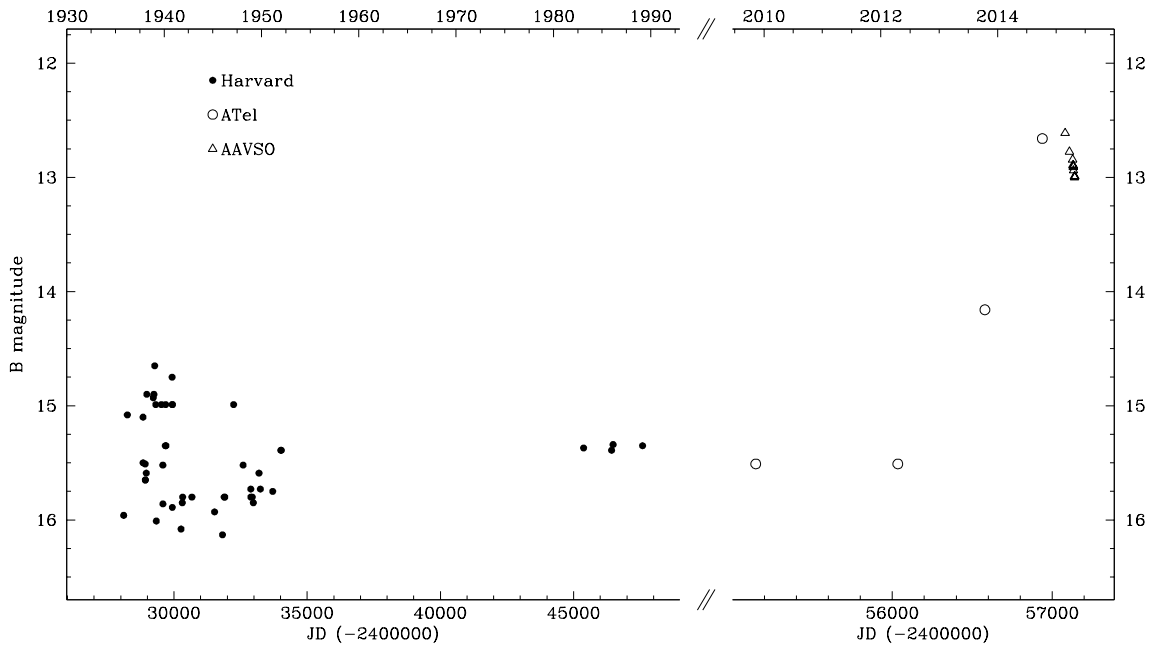}
     \caption{The $B$-band brightness of 2MASS
     J06593158-0405277 on Harvard plates compared with the $B$-band lightcurve
     during the 2014 outburst (see text for details). Note the break
     in the abscissae.}
     \label{fig3}
  \end{figure*}

Historic Harvard plates are almost invariably blue sensitive emulsions
exposed unfiltered through lens astrographs.  For cool objects like 2MASS,
this combination results in a pass-band close to modern $B$ band.  Thus, on
the Harvard plates that we inspected, the brightness of 2MASS was estimated
against the $B$-band sequence of Table~1 and Figure~1.  This, however, does
not necessarely apply to much hotter objects, for which the emission in the
$U$ band is not negligible compared to that through the $B$ band.  For them,
a correction factor has to applied to their brightness estimated on Harvard
plates against a $B$ band sequence (cf.  the template case of the progenitor
of nova KT Eri we have recently investigated on Harvard plates; Jurdana-{\v
S}epi{\'c} et al.  2012, Munari and Dallaporta 2014).  The correction factor
depends on the stellar spectral energy distribution between the $B$ band and
the ultraviolet atmospheric cut-off, the atmospheric and objective lens
transmission over the same wavelength, and the sensitivity as function of
wavelength of the photographic emulsion.

   \begin{table}[!Ht]
     \caption{Upper limits to the $B$ band brightness of 2MASS
              J06593158-0405277 on Harvard plates (this long
              table is published in its entirety in electronic
              form only. A portion is shown here for guidance regarding
              its form and content).}
      \centering
      \includegraphics[width=64mm]{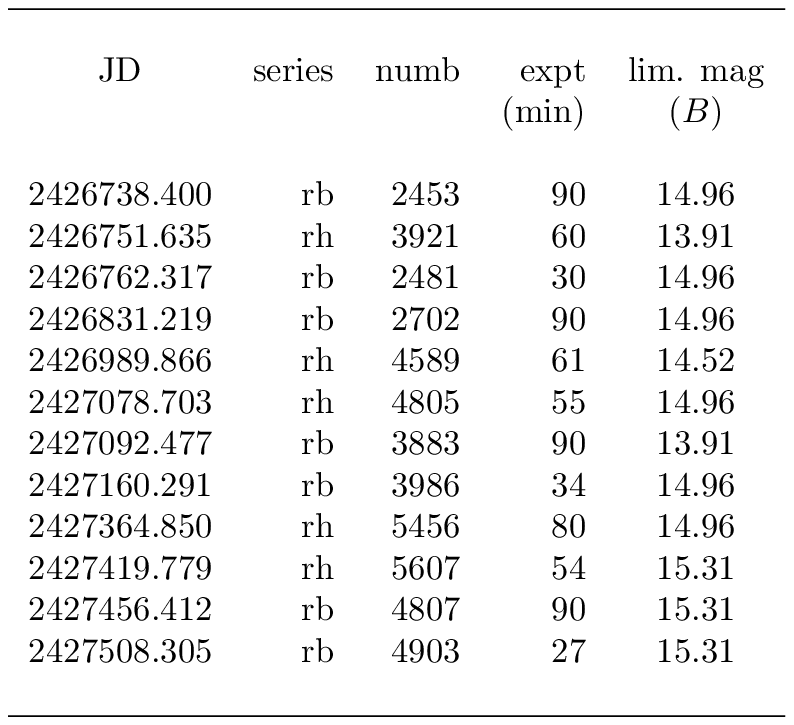}
      \label{tab3}
  \end{table}

Plates potentially covering the 2MASS field were selected for the Harvard
general database based on their plate centres, angular extension and
orientation on the sky.  The plates were manually retrieved from the plate
stack, placed under a high quality monocular lens and the area encompassing
2MASS and the surrounding comparison stars centred on the eyepiece.  A total
of 481 Harvard plates were selected for inspection.  When viewed at the
eyepiece, 103 of them turned out to be unsuited because of a range of
problems (out of focus, too shallow exposure, damaged plate, etc.).  Of the
remaining 378 usable plates (spanning the time interval from 1889-1989 ),
2MASS was fainter than the limiting magnitude on 330 plates, and visible and
measured on 47 of them.  To check upon the accuracy of our measurements,
after a week (to ensure full loss of our memory about them) we remeasured
the brightness of 2MASS on these 47 plates.  The median difference in the
magnitude estimated in these two passes is 0.14 mag, so 0.07 mag from the
mean value, which is a fair estimate of the error associated to our visual
estimates, and in line with the intrinsic accuracy of photographic emulsions
as photometric detectors.  The results are presented in Tables~2 (positive
detections, average of the two independent estimates) and 3 (upper limits to
the brightness of 2MASS based on the faintest star of the photometric
sequence still visible, this long table being available electronic only) and
are plotted in Figures~2 and 3.

We have also inspected the original prints of the {\em Cart du Ciel} all-sky
photographic survey available at the Asiago observatory.  The 2MASS area is
covered by a plate taken on 25 February 1908 at San Fernando Observatory. 
On this plate, star {\em ``m"} of the local photometric sequence is clearly
visible but 2MASS is not, which place an upper limit of 14.5 to the $B$-band
brightness of 2MASS on that date.

\section{Results}

The results of our inspection of Harvard plates are presented in Figure~2,
and fully support the uniqueness of its recent FUOR outburst.  Apart from
the uncovered Menzel's gap in the 50ies (during which photographic sky
patrol was suspended at Harvard), the object has generally remained fainter 
than 14 mag in $B$ band, and when detected the median value for its $B$ band
magnitude has been 15.5.

These results on 2MASS from Harvard plates are compared to photometry of the
current FUOR outburst in Figure~3.  Two sources for the recent data are
considered.  Once the outburst was announced, AAVSO observers acquired data
in the $B$ band (and other optical bands too), and these are plotted in
Figure~3 as daily averages.  They show a decline of about half a magnitude
during the two months they cover before the conjunction with the Sun
prevented further observations.  The second source of recent data are the
discovery and pre-discovery observations by Maehara et al.  (2014) and
Hackstein et al.  (2014).  These observations have been carried out in the
$I_{\rm C}$ band.  To transform them into $B$ band data, we applied a color
index of $B$$-$$I_{\rm C}$=+2.76 as derived from outburst observations by
AAVSO observers. Given the nice agreement with Harvard and AAVSO data in
Figure~3, this transformation looks reasonable. With some degree of
extrapolation, Figure~3 suggests that until late 2012 / early 2013 2MASS
was still in quiescence and that the rise toward maximum took about 1.5
years, a value typical of FUOR (Hartmann and Kenyon 1996).

A final comment is in order about the variability of $\sim$1 mag amplitude
displayed by 2MASS during 1935-1950 (cf Figure~3), which contrasts with the
photometric stability at later epochs and the report by Hackstein et al. 
(2014) of a constancy in brightness within 0.05 mag of 2MASS during
2009-2012.  The variability observed in 1935-1941 is much larger than the
measurement errors (see previous section), and therefore appears real.  It
could either be ascribed to 2MASS itself and its probable T Tau behavior
before the current FUOR outburst, or to the optical companion 2MASS
J06593168-0405224 in blend with our target on all Harvard plates.  Given the
much fainter magnitude of the companion, if this is the one responsible for
the activity recorded during 1935-1950, it must have varied by several
magnitudes in order to affect the combined magnitude by $\sim$1 mag.

\section{Acknowledgements}

We would like to thank Alison Doane, Curator of Astronomical Photographs at
the Harvard College Observatory, for granting access to the Harvard plate
stack, and Stella Kafka for hospitality at AAVSO Headquarters.  This work
was supported in part by the University of Rijeka under the project number
13.12.1.3.03.  and by the Croatian Science Foundation under the project 6212
Solar and Stellar Variability.

\end{document}